\documentclass[english,aps,preprint]{revtex4}
\usepackage[T1]{fontenc}
\usepackage[latin9]{inputenc}
\usepackage{array}
\usepackage{multirow}
\usepackage{amsmath}
\usepackage{graphicx}

\makeatletter

\providecommand{\tabularnewline}{\\}

\@ifundefined{textcolor}{}
{%
 \definecolor{BLACK}{gray}{0}
 \definecolor{WHITE}{gray}{1}
 \definecolor{RED}{rgb}{1,0,0}
 \definecolor{GREEN}{rgb}{0,1,0}
 \definecolor{BLUE}{rgb}{0,0,1}
 \definecolor{CYAN}{cmyk}{1,0,0,0}
 \definecolor{MAGENTA}{cmyk}{0,1,0,0}
 \definecolor{YELLOW}{cmyk}{0,0,1,0}
}

\makeatother

\usepackage{babel}
\begin{document}

\title{Single production of the excited electrons at the future FCC-based
lepton-hadron colliders}

\author{A. Caliskan}

\email{acaliskan@gumushane.edu.tr}

\affiliation{Gümü\c{s}hane University, Faculty of Engineering and Natural Sciences,
Department of Physics Engineering, 29100, Gümü\c{s}hane, Turkey}

\author{S. O. Kara}

\email{sokara@ohu.edu.tr}

\affiliation{Omer Halisdemir University, Bor Vocational School, 51240, Nigde,
Turkey}
\begin{abstract}
We study composite electron production at the FCC-based three electron-proton
colliders with the center-of-mass energies of $3.46$, $10$ and $31.6$
TeV. For the signal process of $ep\rightarrow e^{\star}X\rightarrow e\gamma X$
, the production cross-sections and decay widhts of the excited electrons
have been calculated. The differences of some kinematical quantities
of the final state particles between the signal and background have
been analyzed. For this purpose, transverse momentum and pseudorapidity
distributions of electron and photon have been obtained and the kinematical
cuts for discovery of the excited electrons have been assigned. We
have finally determined the mass limits of excited electrons for observation
and discovery by applying these cuts. It is shown that the mass limit
for discovery obtained from the collider with $\sqrt{s}=31.6$ TeV
(called PWFA-LC$\otimes$FCC) is $22.3$ TeV for the integrated luminosity
$L_{int}=10$ $fb^{-1}$.
\end{abstract}
\maketitle

\section{introduction}

The Standard Model (SM) of particle physics provides a successful
description of the properties of electromagnetic, weak and strong
interactions of the elementary particles. It also shows a great concordance
with the all experiments carried out up to date, and has been finally
reached its last estimation on Higgs particle by the ATLAS and the
CMS collaborations, which both announced the discovery of this particle
in 2012 \cite{1,2}. Since there are some phenomena that the SM does
not explain, such as large number of elementary particles, quark-lepton
symmetry and family replication, it is not a fundamental theory in
particle physics, but is supposed to be effective model of a more
comprehensive theory. To find a solution to these deficiencies in
the SM, a lot of theory beyond the SM (BSM) have been therefore proposed
until now, such as supersymmetry (SUSY), grand unified theories (GUTs),
extra dimensions and compositeness. The compositeness is one of the
most important BSM theories, because it explains the subjects of the
family replication, the inflation of elementary particles and the
quark-lepton symmetry in the best manner, introducing the more fundamental
matter and antimatter constituents in which the leptons and quarks
have a substructure called preons \cite{3}.

The lepton and quark compositeness were first proposed at the end
of 1970s \cite{4,5,6,7}, and many preonic models have been discussed
until today, such as haplon model \cite{8,9}, rishon model \cite{10,11},
and so on. Many new types of particles have been suggested in the
framework of the preonic models, for example, diquarks, excited fermions,
leptoquarks, color sextet quarks, color sextet leptons, dileptons
and leptogluons. As a results of the compositeness, new interactions
among the fermions should occur at the scale of the constituent binding
energies. This energy scale is a characteristical parameter of the
composite models, and called the compositeness scale, $\varLambda$.

If the known fermions have a substructure they should be regarded
as the ground state to a rich and heavier spectrum of the excited
states. Excited leptons and quarks are predicted by the composite
models. We have interested in the excited electrons with spin-1/2
in this paper, as a continuation of our previous works performed for
the excited muons \cite{12} and neutrinos \cite{13} at the FCC-based
lepton-hadron colliders. In addition to these studies there are also
many important phenomenological studies carried out recently in the
literature for the excited leptons \cite{14,15,16,17-1,18-1,19-1,20-1}
and quarks \cite{21-1,22-1}.

Even though there is no evidence for the excited leptons in the experimental
studies performed in HERA \cite{23-1}, Tevatron \cite{24-1}, ATLAS
\cite{25-1} and CMS \cite{26-1}, the colliders with higher center-of-mass
energy and luminosity, planned to be installed in the future, will
continue to be hope for their discovery. A possible discovery of the
any excited fermion will be a direct proof of the lepton and quark
compositeness. The most recent experimental results on the excited
electron mass are provided by the OPAL and the ATLAS collaborations
\cite{27-1}. The mass exclusion limits of the excited electrons are
$m_{e^{\star}}$ > 103.2 GeV for pair production ($e^{+}$$e^{-}$$\rightarrow$$e^{\star}$$e^{\star}$)
and $m_{e^{\star}}$ > 3000 GeV for single production, assuming $f=f'=1$
and $\varLambda=m_{e^{\star}}$. 

We have analyzed the production potential of the excited electrons
at the future electron-proton colliders. We present the main parameters
of the FCC-based ep colliders in the section 2, the interaction Lagrangian
responsible for the gauge interactions of the excited electrons, their
decay widths and the cross-sections in the section 3, and the signal
and background analysis in the section 4. Finally we summarized the
all results in the last section.

\section{THE FCC-BASED ELECTRON-PROTON COLLIDERS}

The Large Hadron Collider (LHC) is the most powerful energy-frontier
hadron collider ever constructed. The LHC is presently in operation
at the CERN, and will continue to work until the middle of 2030s in
the framework of its high-luminosity upgrade programme. The international
Future Circular Collider (FCC) study \cite{28-1} has been launched
in 2010-2013 at the CERN, and supported by European Union within the
Horizon 2020 Framework for Research and Innovation, as a next-generation
collider for the post-LHC era. Its main goal is to construct an energy-frontier
hadron collider with a center-of-mass energy of the order of 100 TeV
in a new $\sim$ 100 km tunnel near the Geneva. The FCC will have
almost 4 times bigger in circumferences, and nearly 7 times higher
center-of-mass energy than ones of the LHC. The FCC hadron collider
(FCC-hh) could enable us to search for the physics of the BSM theories
at the highest energies. The Conceptual Design Report (CDR) of the
FCC is expected to be issued in 2018.

The FCC project also includes the design of a 90-400 GeV high luminosity
electron-positron collider (FCC-ee or TLEP) \cite{29-1}, which could
be installed in the same tunnel, to search the top quark, W, Z and
Higgs bosons, as an intermediate step of the project. Construction
of a such collider in the same tunnel will give us the opportunity
to collide the proton beam with the electron beam, which known as
FCC-he option. The energy of the electrons in the electron ring of
the TLEP is limited because of the large synchrotron radiation. Therefore,
to reach higher energy electron beam the establishment of a linear
electron accelerator tangential to the FCC ring has been recently
proposed in \cite{30-1}, using the parameters of the known linear
electron collider projects, namely ILC (International Linear Collider)
\cite{31-1} and PWFA-LC (Plasma Wake Field Accelerator - Linear Collider)
\cite{32-1}. Thus, the FCC-based many electron-proton collider options
have been obtained when we also consider the nominal energy values,
that can be upgraded, of the ILC and the PWFA-LC.

\begin{table}

\caption{Main parameter of the FCC-based ep colliders.}

\begin{centering}
\begin{tabular}{|c|c|c|c|}
\hline 
Colliders & $E_{e}$(TeV) & CM Energy (TeV) & $L_{int}$($fb^{-1}$per year)\tabularnewline
\hline 
\hline 
ERL60$\otimes$FCC & $0.06$ & $3.46$ & $100$\tabularnewline
\hline 
ILC$\otimes$FCC & $0.5$ & $10$ & $10-100$\tabularnewline
\hline 
PWFA-LC$\otimes$FCC & $5$ & $31.6$ & $1-10$\tabularnewline
\hline 
\end{tabular}
\par\end{centering}

\end{table}

In the numerical calculations, we have used the parameters of the
FCC-based electron-proton colliders shown in the Table 1, in which
there are three collider options that have the different center-of-mass
energies. In here, the ERL60 denotes the Energy Recovery Linac with
the electron energy of 60 GeV, which had been chosen as the main option
for the LHeC \cite{33-1}. The same ERL can be used for the FCC-based
ep collider \cite{34-1}.

\section{Interact\i on lagrang\i an, decay w\i dths and cross-sect\i ons}

For the interaction of a spin-1/2 excited lepton with the SM leptons
and a gauge boson we have used the following SU(2)xU(1) invariant
Lagrangian \cite{35-1,36,37,38},

\begin{center}
\begin{equation}
L=\frac{1}{2\Lambda}\bar{l_{R}^{\star}}\sigma^{\mu\nu}[fg\frac{\overrightarrow{\tau}}{2}.\overrightarrow{W}_{\mu\nu}+f'g'\frac{Y}{2}B_{\mu\nu}]l_{L}+h.c.,
\end{equation}

\par\end{center}

where $l$ and $l^{\star}$ represent the SM lepton and the excited
lepton, respectively, $\Lambda$ is the compositeness scale, $g$
and $g'$ are the gauge couplings, $\overrightarrow{W}_{\mu\nu}$
and $B_{\mu\nu}$ are the field strength tensors, $f$ and $f'$ are
the scaling factors for the gauge couplings, Y is hypercharge, $\sigma^{\mu\nu}=i(\gamma^{\mu}\gamma^{\nu}-\gamma^{\nu}\gamma^{\mu})/2$
where $\gamma^{\mu}$ are the Dirac matrices, and $\overrightarrow{\tau}$
denotes the Pauli matrices.

For the excited electrons, three decay channels are possible: $\gamma$-channel
($e^{\star}\rightarrow e\gamma$), Z-channel ($e^{\star}\rightarrow eZ$)
and W-channel ($e^{\star}\rightarrow eW$). We have chosen the electromagnetic
decay mode ($\gamma$-channel) for this study, because of the easy
detection of this channel compared to the others.

Ignoring the SM electron mass, the decay widths of the excited electrons
are obtained as, 

\begin{equation}
\varGamma(l^{\star}\rightarrow lV)=\frac{\alpha m^{\star3}}{4\Lambda^{2}}f_{V}^{2}(1-\frac{m_{V}^{2}}{m^{\star2}})^{2}(1+\frac{m_{V}^{2}}{2m^{\star2}}),
\end{equation}

where $m^{\star}$ is the mass of the excited electron, $f_{V}$ is
the new electroweak coupling parameter corresponding to the gauge
boson V, where V=W, Z, $\gamma$, and $f_{\gamma}=-(f+f')/2$, $f_{Z}=(-f\cot\theta_{W}+f\tan\theta_{W})/2$,
$f_{W}=(f/\sqrt{2}\sin\theta_{W})$, where $\theta_{W}$ is the weak
mixing angle, and $m_{V}$ is the mass of the gauge boson.

\begin{figure}
\begin{centering}
\includegraphics[scale=0.8]{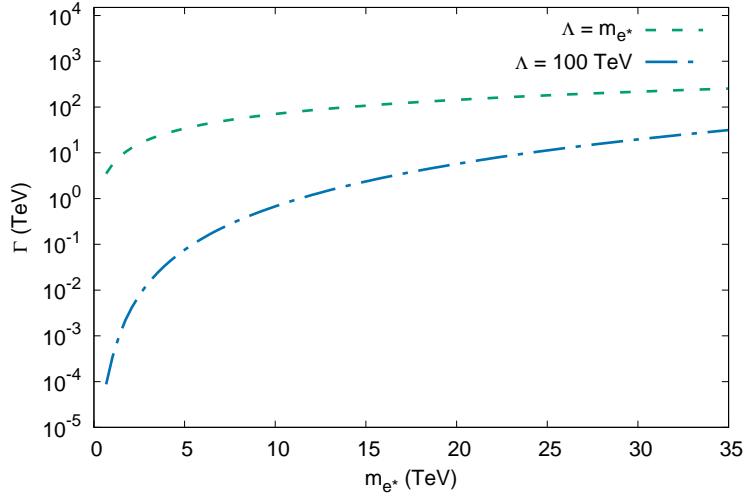}
\par\end{centering}

\caption{The total decay widths of the excited electron for the scale $\varLambda=m_{e^{\star}}$
and $\varLambda=100$ TeV, assuming the coupling $f=f'=1$.}

\end{figure}

For the numerical calculations we have implemented the excited electron
interaction vertices into the CALCHEP \cite{39}, which is a high-energy
simulation programme. We show the total decay widths of the excited
electron in Fig.1, for $\varLambda=m_{e^{\star}}$ and $\varLambda=100$
TeV, which are commonly used for the new physics scale. Figure 2 presents
the total cross-sections for the excited electron production at the
three electron-proton colliders, which are ERL60$\otimes$FCC, ILC$\otimes$FCC
and PWFA-LC$\otimes$FCC, using the CALCHEP program with the CTEQ6L
parton distribution functions \cite{40}. It is clearly seen from
this figure that the excited electrons have adequately high cross-sections
for $\varLambda=m_{e^{\star}}$ and $\varLambda=100$ TeV, at the
high mass values of the excited electron.

The four fermion contact interactions can also contribute to the production
of the excited electrons, besides the gauge interactions. It is particularly
well known that the contact interactions are more dominant at proton-antiproton
colliders. In a recent analysis (see ref.\cite{19-1}), it is shown
that at the LHC energies the production cross-sections of the excited
leptons for the contact interaction are higher than ones of the gauge
interactions. In this study, for the electron-proton type colliders
only gauge interaction mechanism has been taken into account, but
the contribution of contact interaction can not be ignored. Therefore,
the contact interaction version of this work will be addressed in
a future study.

\begin{figure}
\begin{centering}
\includegraphics[scale=0.6]{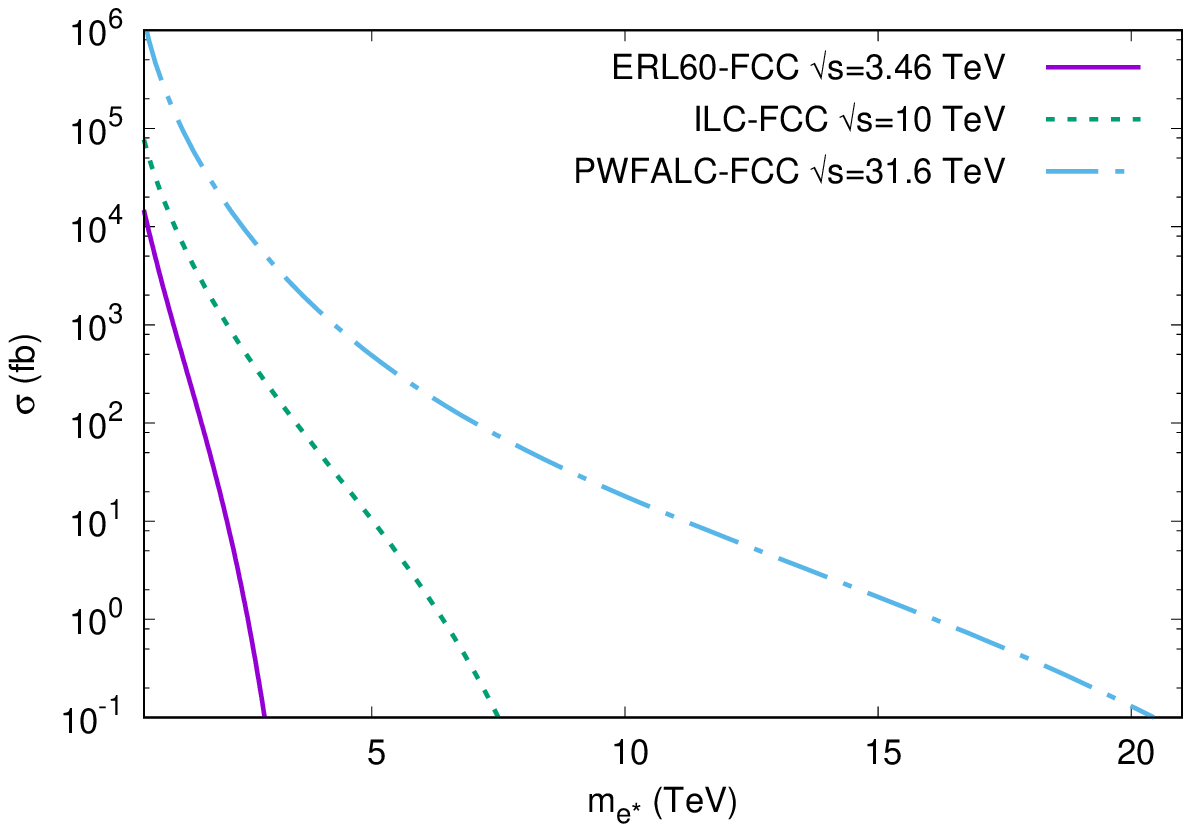}\includegraphics[scale=0.6]{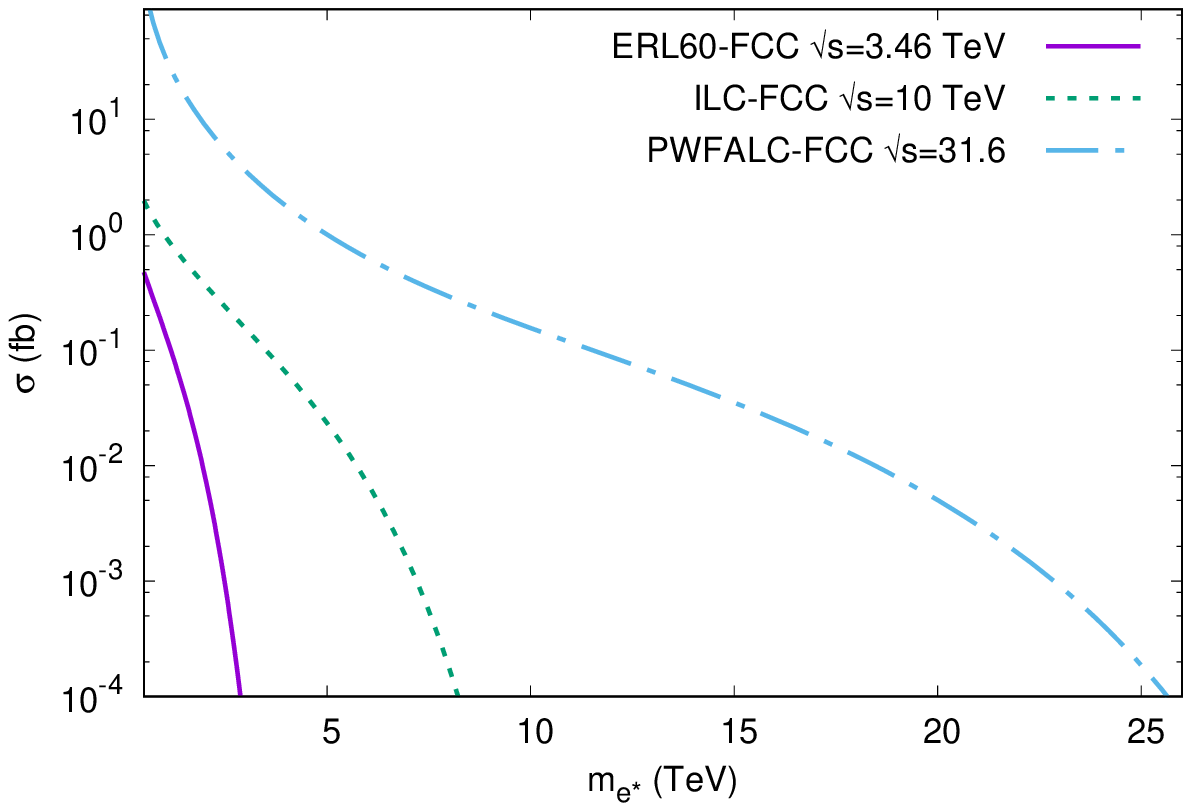}
\par\end{centering}

\caption{The total cross-section values of the excited electrons with respect
to its mass at the ep colliders with various center-of-mass energies
for $\varLambda=m_{e^{\star}}$(left) and $\varLambda=100$ TeV (right),
assuming the coupling $f=f'=1$.}

\end{figure}

\section{SIGNAL AND BACKGROUND ANALYSIS}

The FCC-based electron-proton colliders will allow us to explore the
excited electrons via $ep\rightarrow e^{\star}X$ process with subsequent
decays of the excited electrons into an electron and photon. Thus,
our signal process is $ep\rightarrow e\gamma X$, and subprocesses
are $eq(\overline{q})\rightarrow e\gamma q(\overline{q})$, while
the background process is $ep\rightarrow e,\gamma,j$ through $\gamma$
and $Z$ exchange, where $j$ represents jets which are composed of
quarks ($u,\overline{u},d,\overline{d},c,\overline{c},s,\overline{s},b,$$\overline{b}$).
In this section we discuss the differences of some kinematical quantities
between the signal and the background to determine appropriate kinematical
cuts for discovery of the excited electrons. This analysis is at the
parton level since for a such collider an appropriate detector has
not been designed yet. Figure 3 shows angular distributions and transverse
momentum distributions of the final state particles, electron and
photon, from the ERL60-FCC collider. The pseudorapidity distributions
(top-left and bottom-left) of the signal are peaked almost at $\eta=[-2,-3]$
interval for a given parameter values ($m_{e^{\star}}=1000,2000,3000$
GeV and $\varLambda=m_{e^{\star}}$) of both particles. Since pseudorapidity
is defined as $\eta=-\ln\tan(\theta/2)$, where $\theta$ is polar
angle, it is understand that electrons and photons are of backward,
consequently the excited electrons are produced in the backward direction.
Also, the signal and background are very well separated for the all
mass values. To drastically reduce the background we have applied
a cut on the pseudorapidity as $-5<\eta^{e}<-1$ and $-5<\eta^{\gamma}<-1.5$.

As for the transverse momentum distributions (top-right and bottom-right),
it is clearly seen that the signal and background distributions are
very well separated from each other for both particles. So, we easly
cut off the most of the background by applying a cut at 250 GeV (for
electron) and at 200 GeV (for photon).

\begin{figure}
\begin{centering}
\includegraphics[scale=0.42]{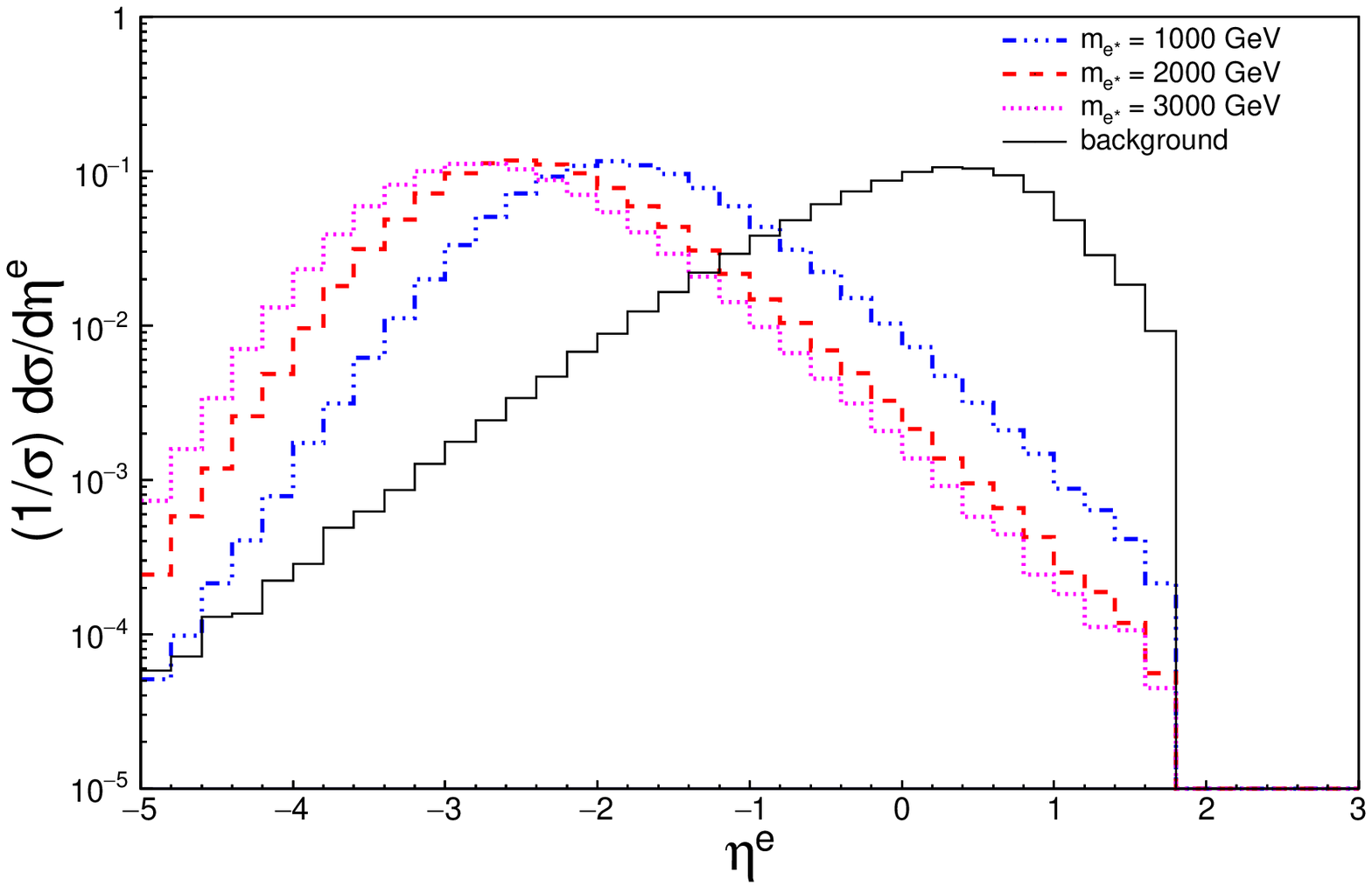}\includegraphics[scale=0.42]{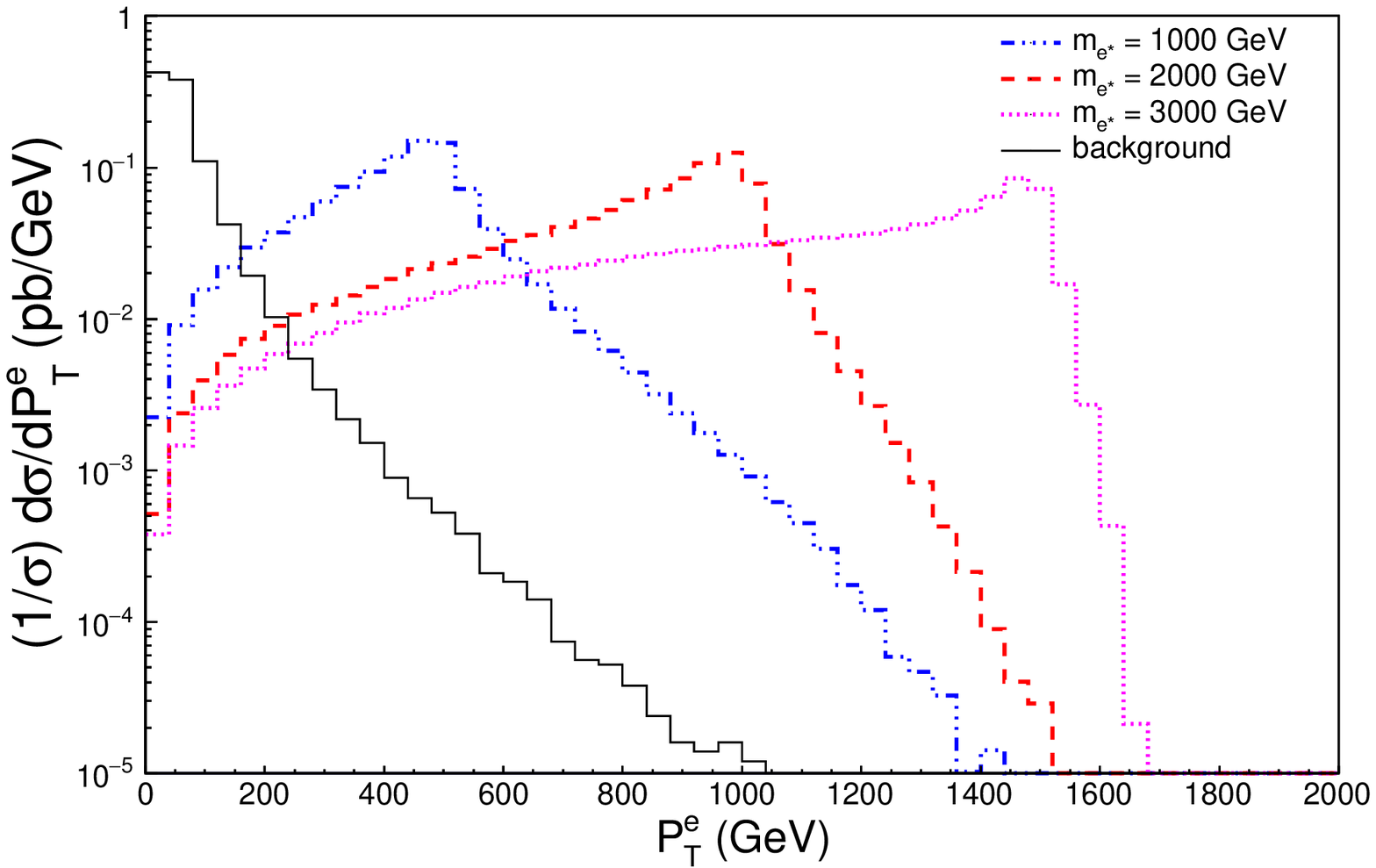}
\par\end{centering}

\begin{centering}
\includegraphics[scale=0.42]{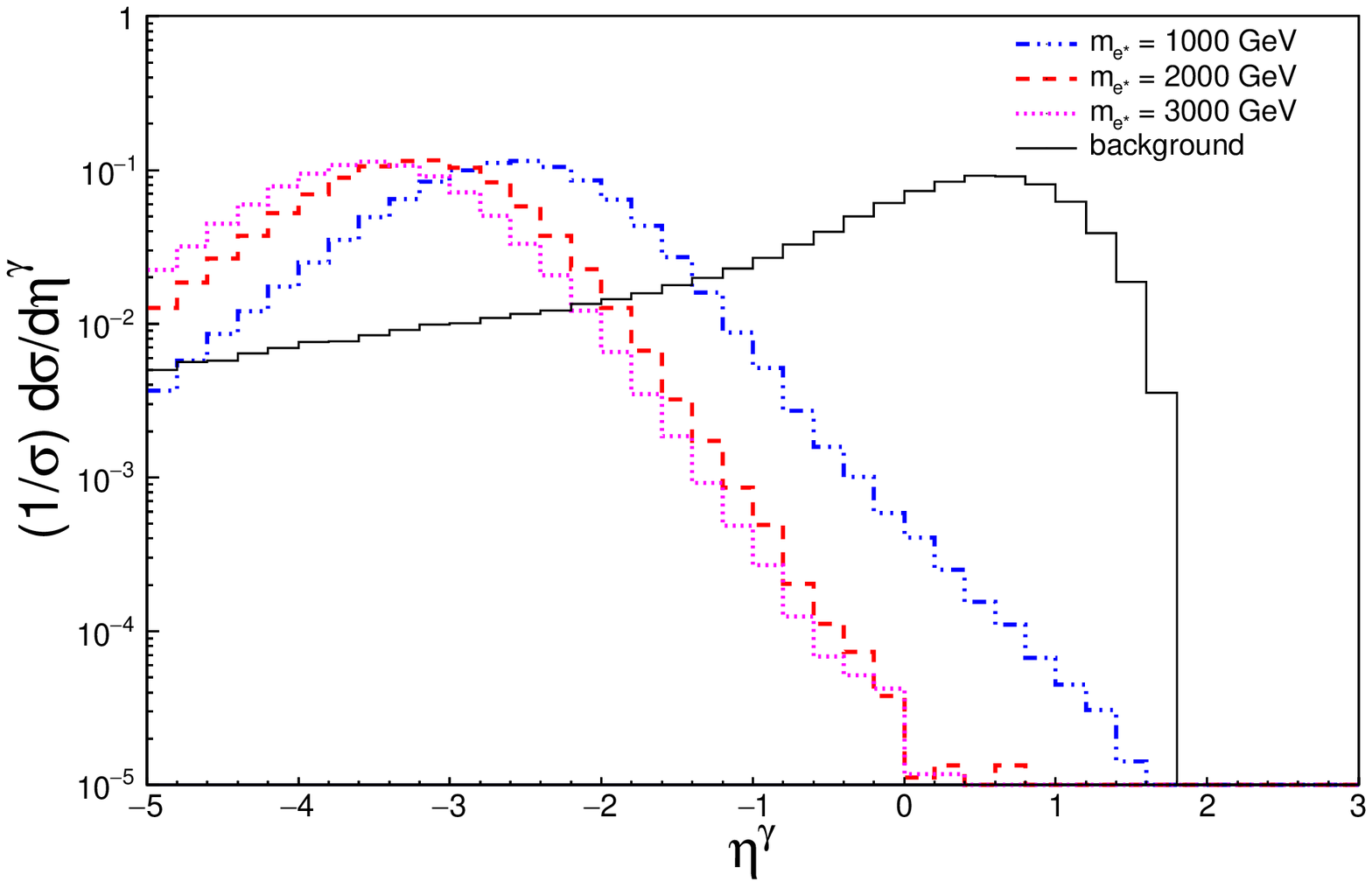}\includegraphics[scale=0.42]{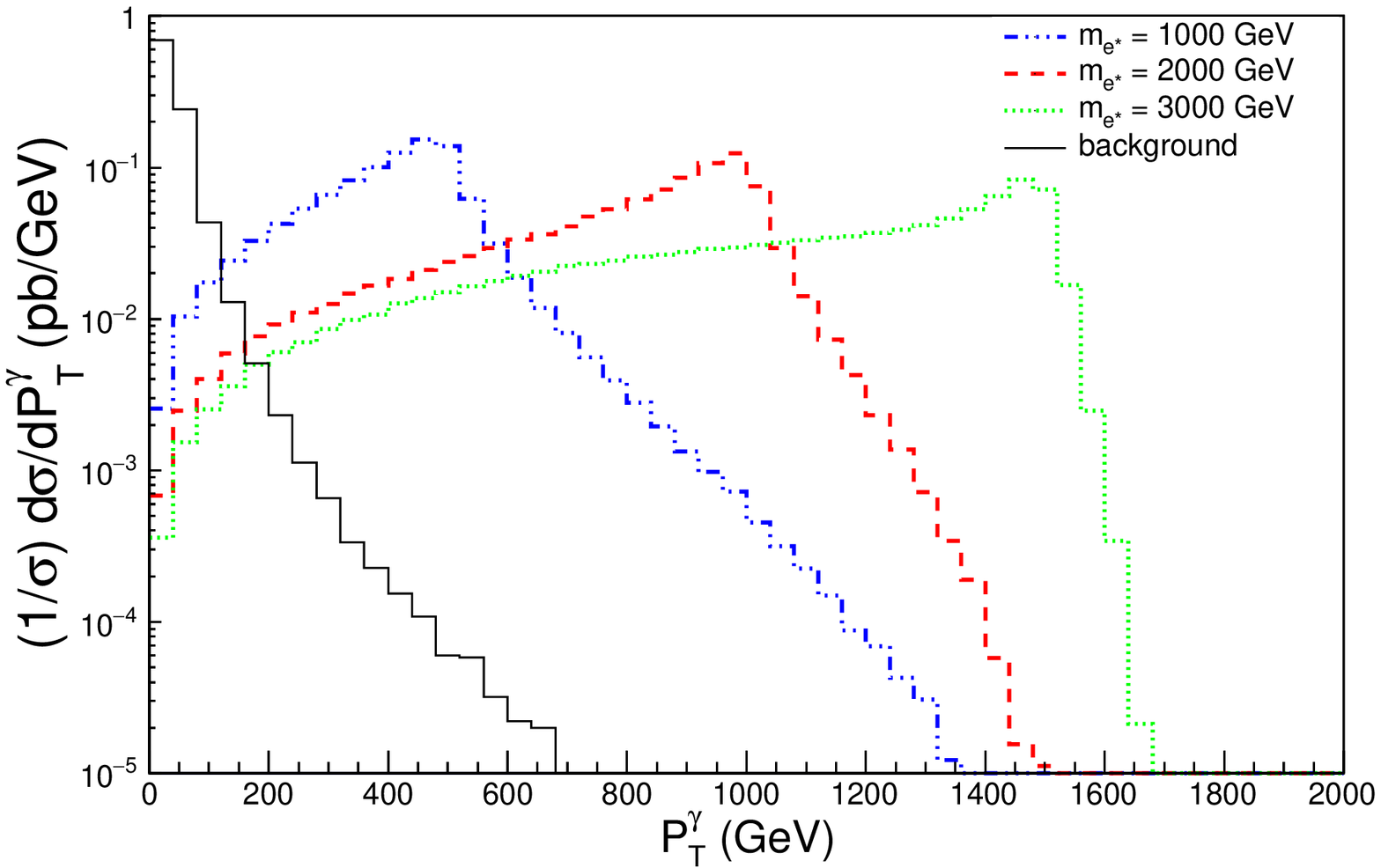}
\par\end{centering}

\centering{}\caption{The normalized pseudorapidity distributions of electron (top-left)
and photon (bottom-left), and normalized tranverse momentum distributions
of electron (top-right) and photon (bottom-right) for $f=f'=1$ and
$\varLambda=m_{e^{\star}}$ at the ERL60$\otimes$FCC collider.}
\end{figure}

The kinematical distributions of the final state particles for the
ILC-FCC collider are shown in the Figure 4. As seen from pseudorapidity
distributions (top-left and bottom-left), the signals are peaked at
the negative region, as for the ERL-FCC. So, the excited electrons
are produced in the backward direction compared to beam axis. We have
easly assigned the kinematical cuts, as $-4<\eta^{e}<0.5$, $-3.5<\eta^{\gamma}<-0.5$,
$p_{T}^{e,\gamma}>600$ GeV, on the pseudorapidity and transverse
momentum distributions for both particles, because the signal and
background are separated very well from each other for the all distributions
in Figure 4. With these cuts the background is substancially suppressed
whereas the signal remains almost unchanged.

\begin{figure}
\includegraphics[scale=0.42]{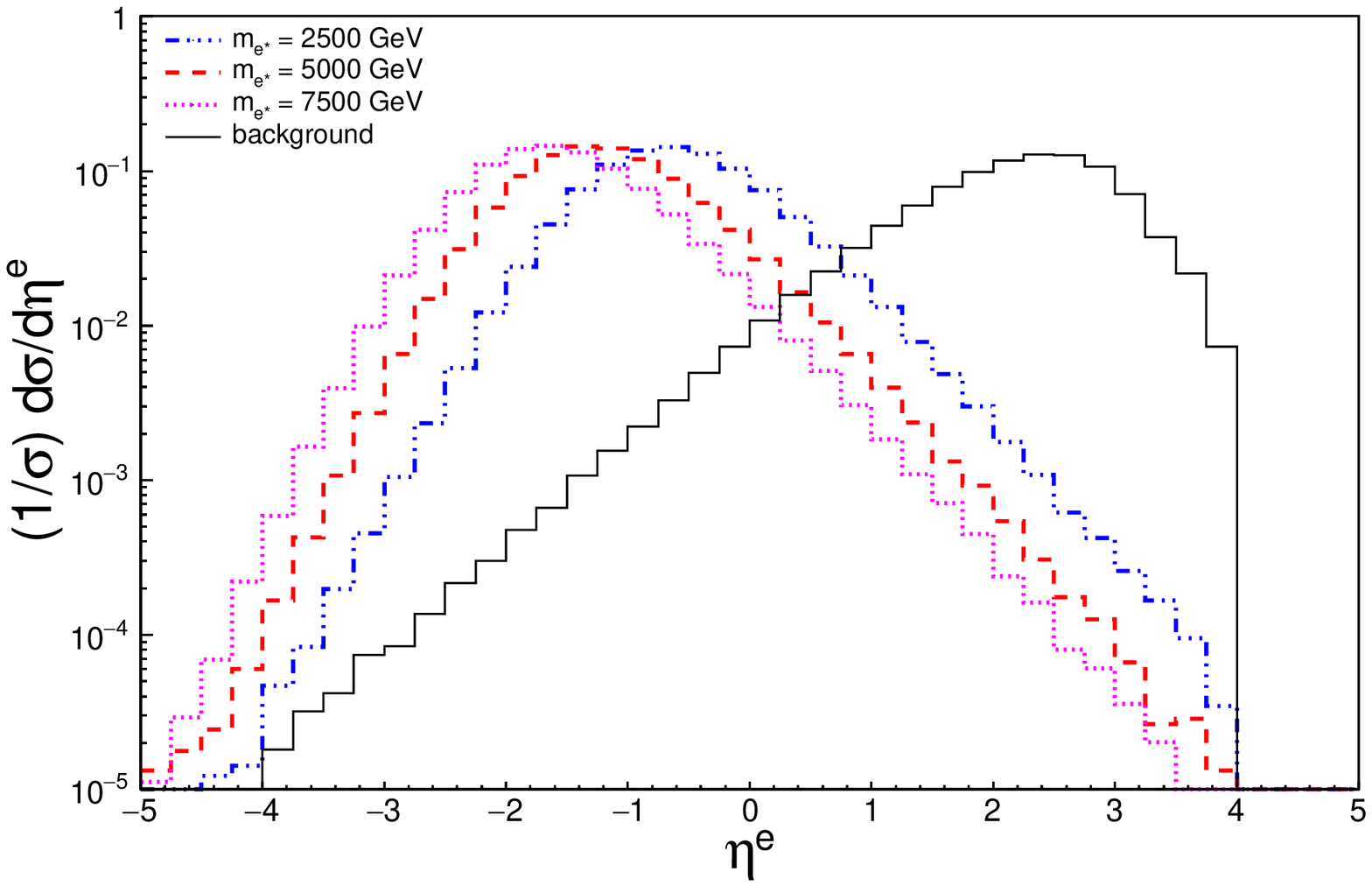}\includegraphics[scale=0.42]{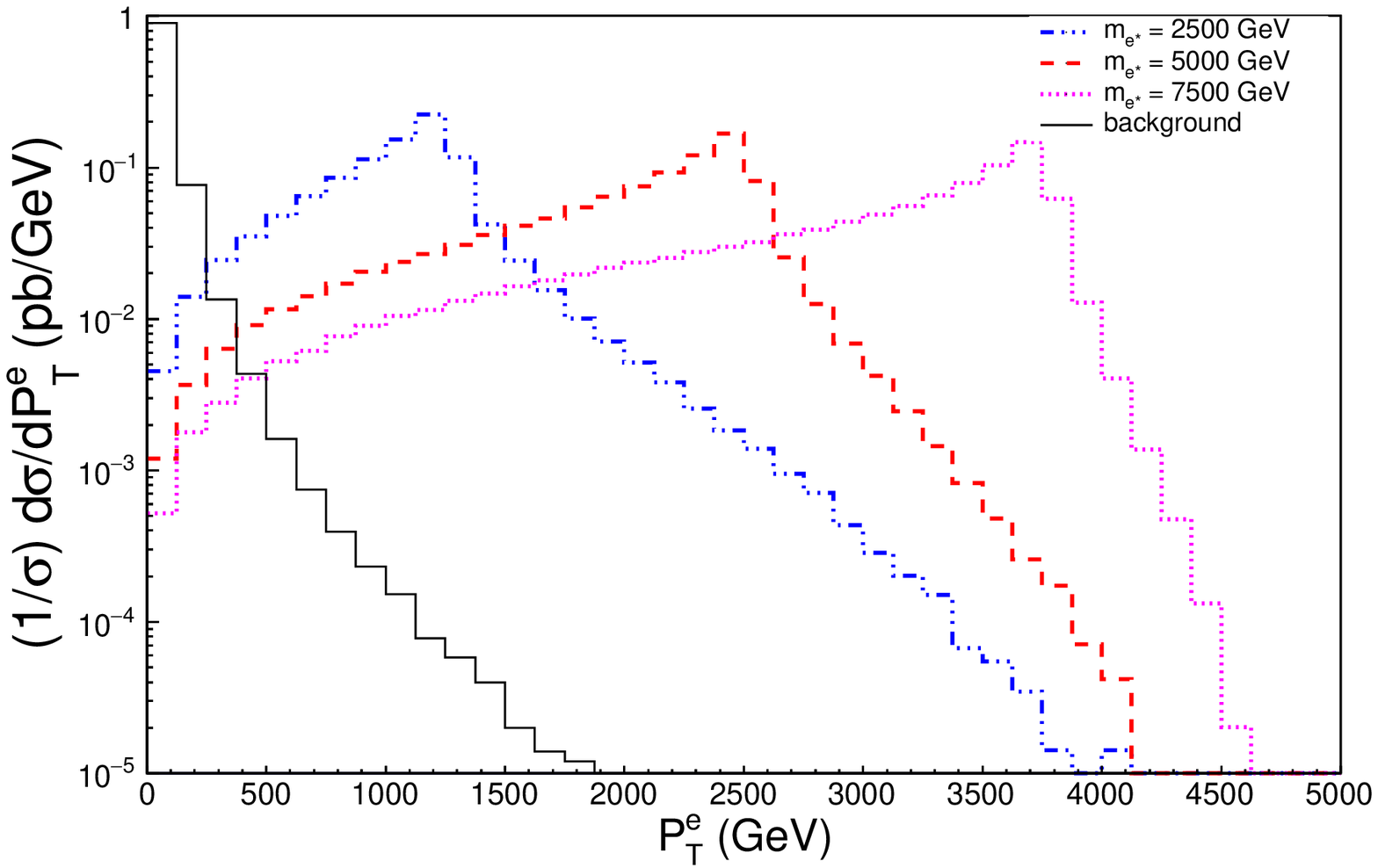}

\includegraphics[scale=0.42]{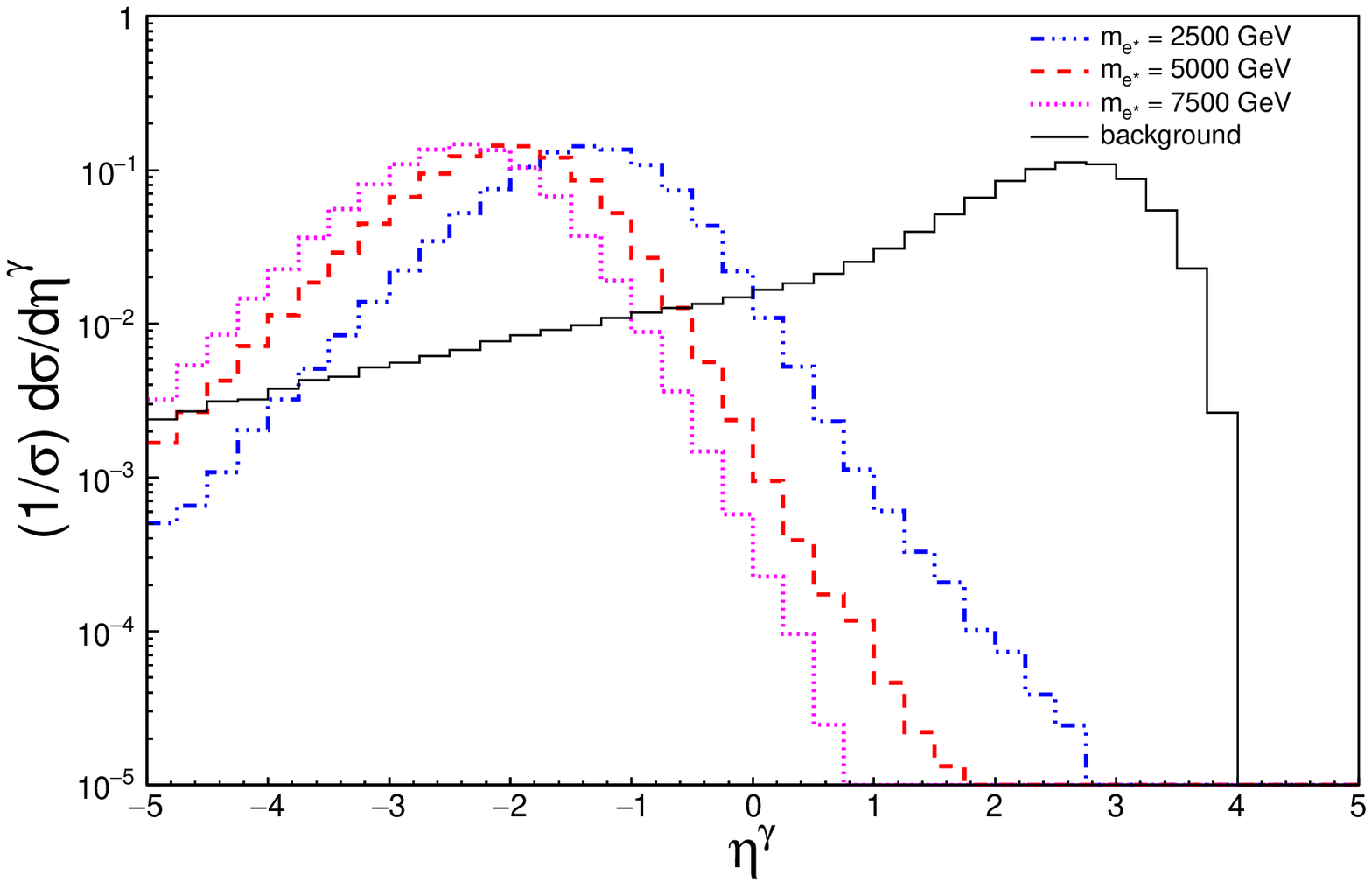}\includegraphics[scale=0.42]{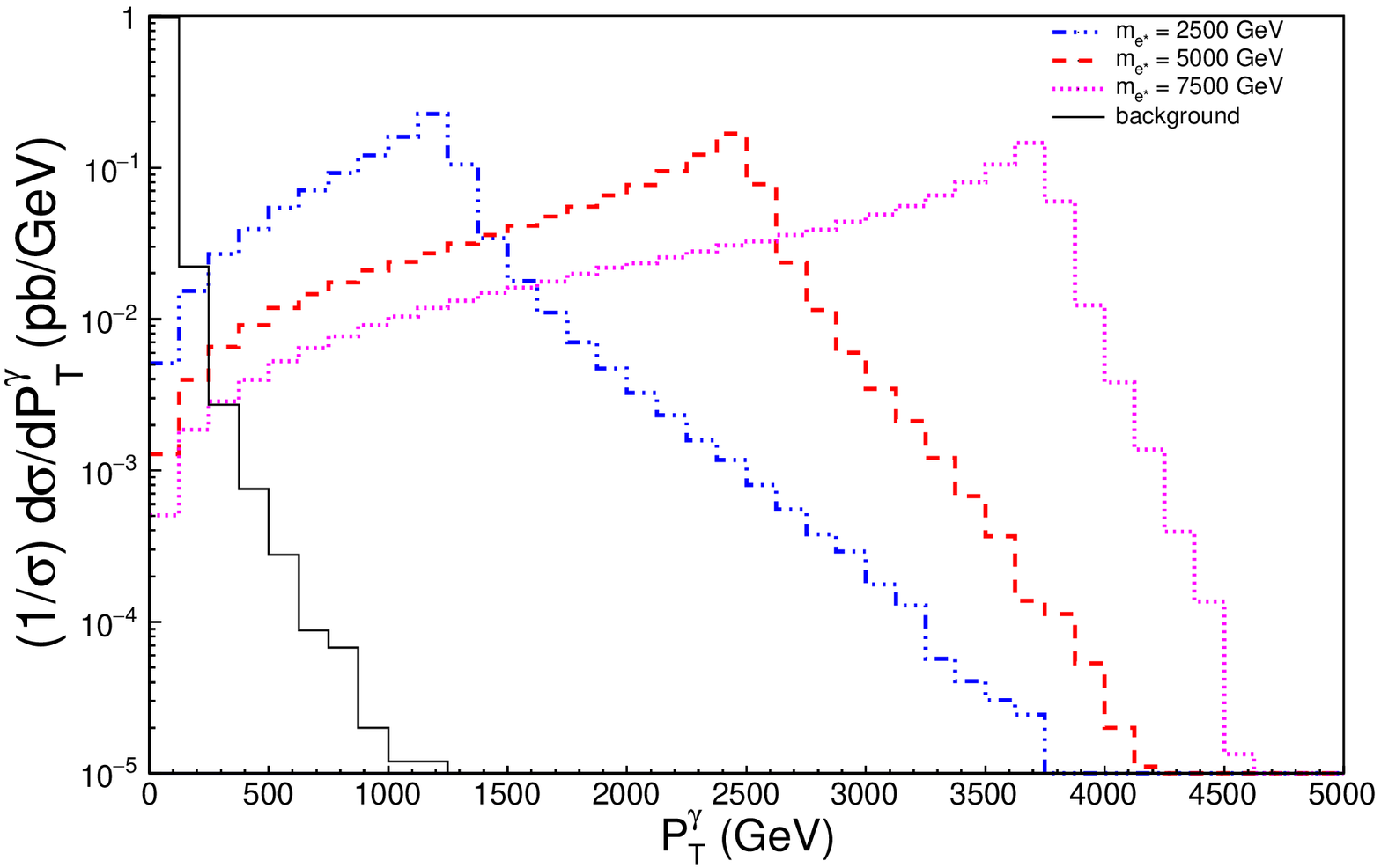}

\caption{The normalized pseudorapidity distributions of electron (top-left)
and photon (bottom-left), and normalized tranverse momentum distributions
of electron (top-right) and photon (bottom-right) for $f=f'=1$ and
$\varLambda=m_{e^{\star}}$ at the ILC$\otimes$FCC collider.}
\end{figure}

The pseudorapidity distributions of the electron and photon for PWFALC-FCC
collider, as seen from Figure 5 (top-left and bottom-left), are slightly
different from those of the previous colliders. The angular distribution
for $m_{e^{\star}}=3000$ GeV are peaked in the positive region for
both final state particles. Thus, the excited electrons with the small
masses like 300 GeV are produced in the forward direction. The kinematical
discovery cuts of this collider are determined as $-3<\eta^{e}<2.5$,
$-2.5<\eta^{\gamma}<2$, $p_{T}^{e,\gamma}>800$ GeV.

\begin{figure}
\includegraphics[scale=0.42]{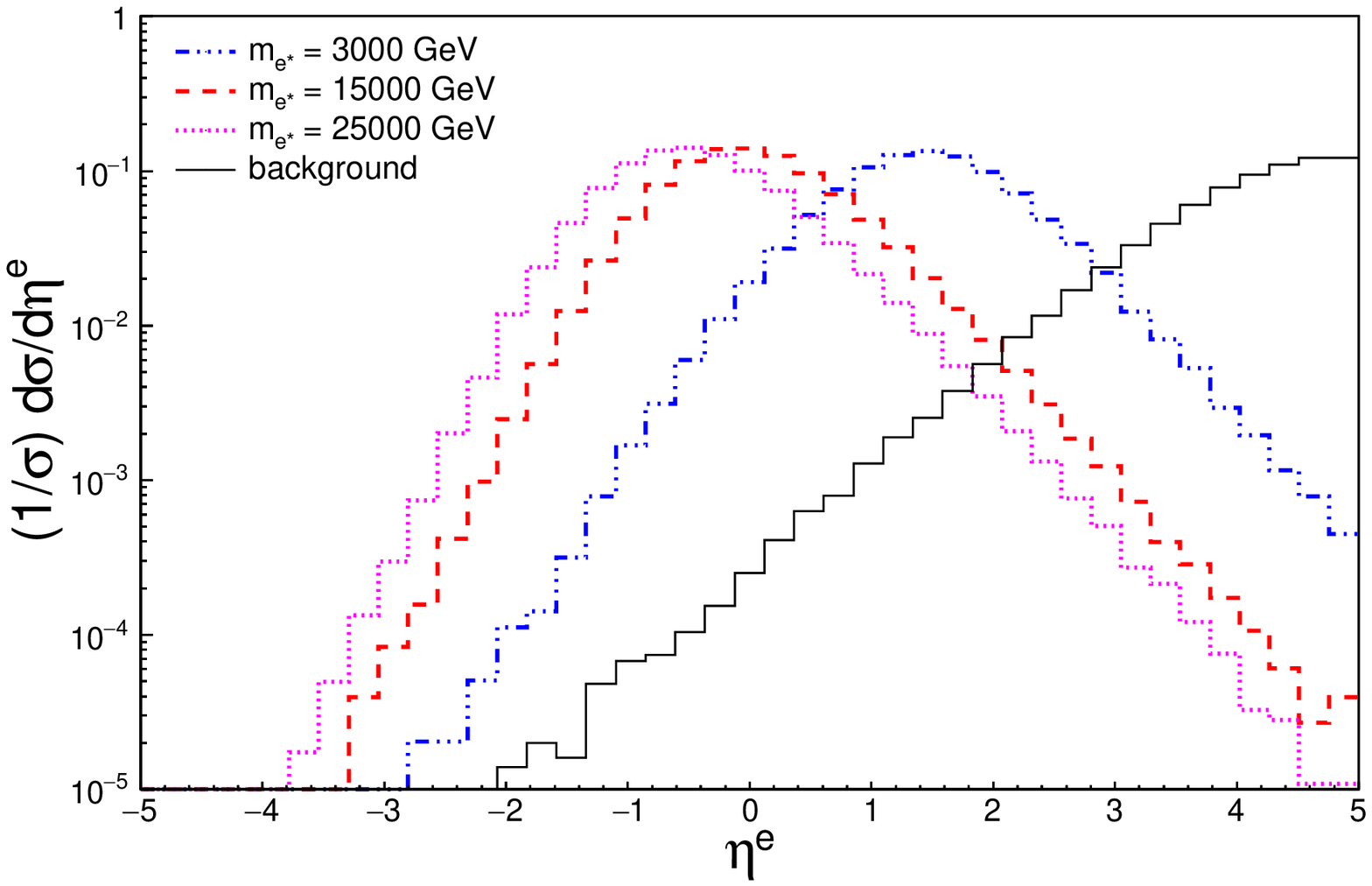}\includegraphics[scale=0.42]{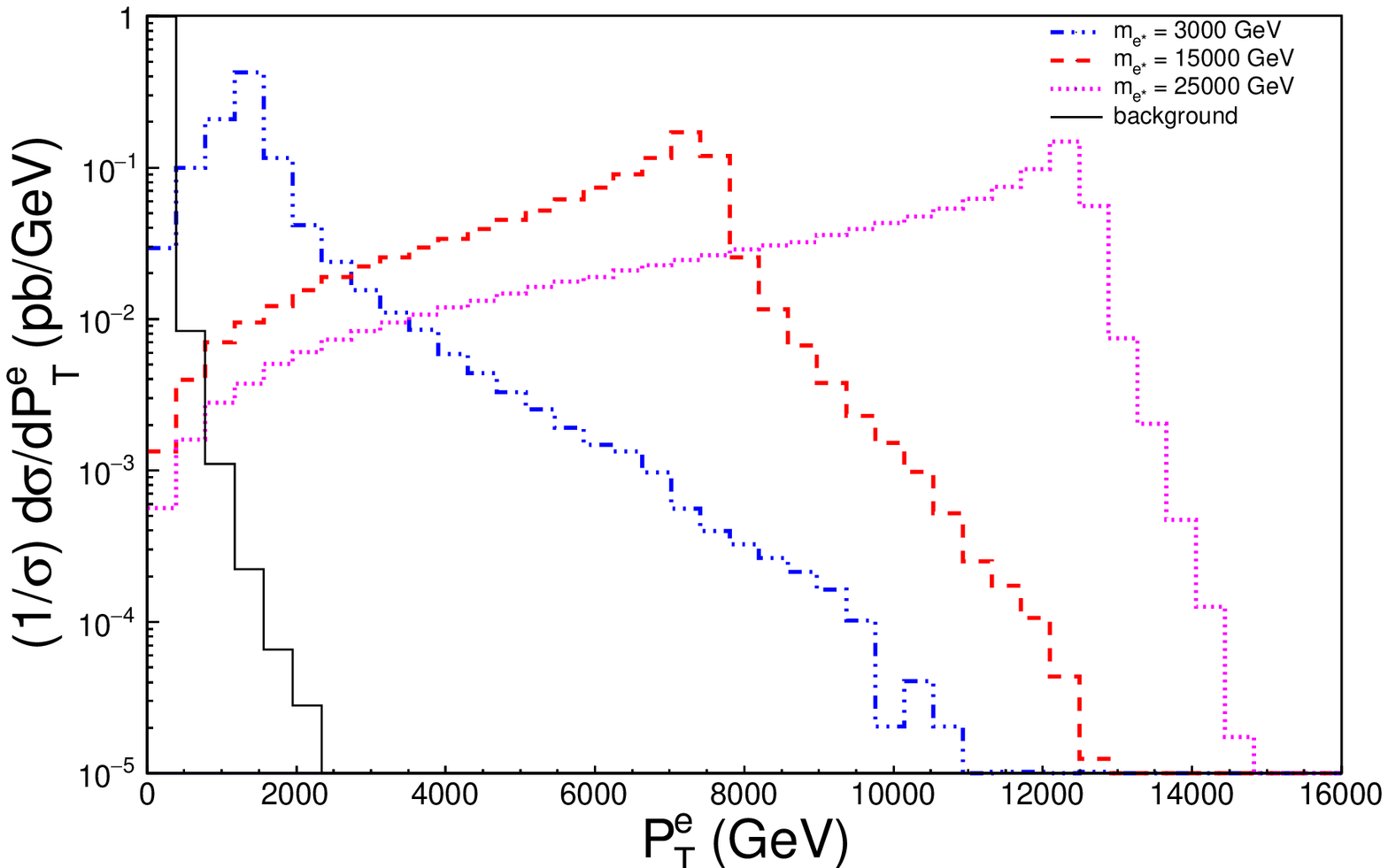}

\includegraphics[scale=0.42]{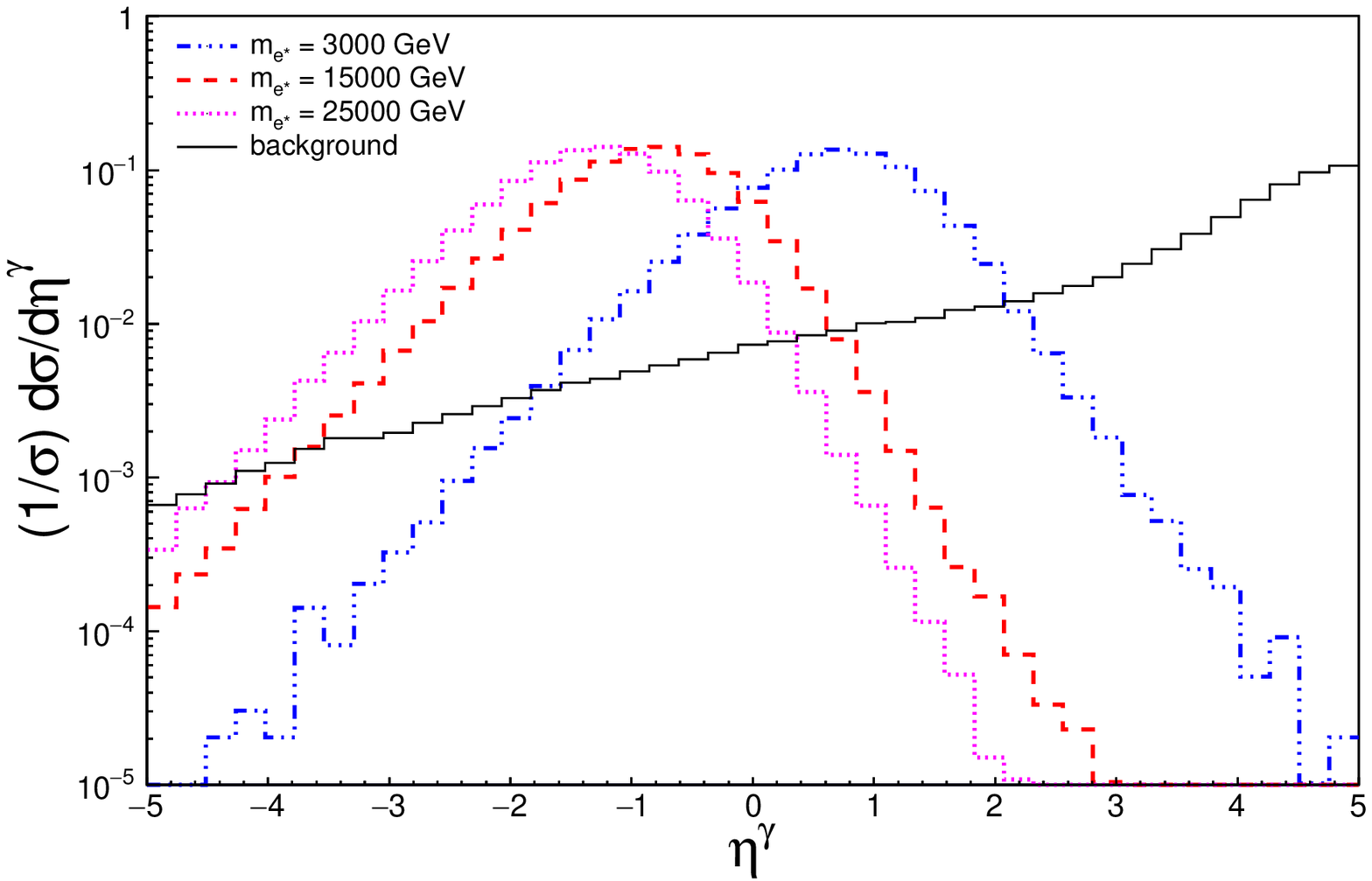}\includegraphics[scale=0.42]{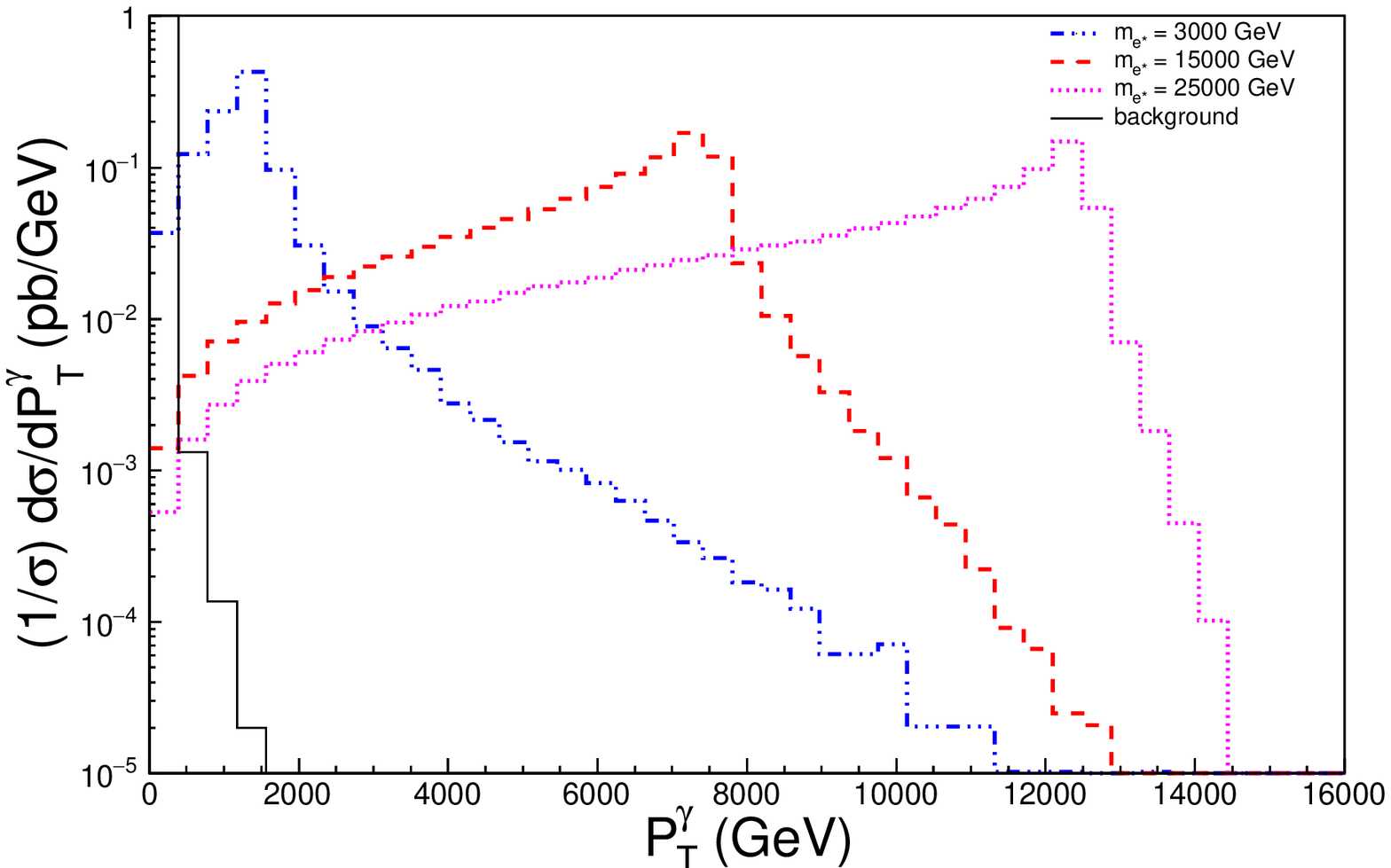}

\caption{The normalized pseudorapidity distributions of electron (top-left)
and photon (bottom-left), and normalized tranverse momentum distributions
of electron (top-right) and photon (bottom-right) for $f=f'=1$ and
$\varLambda=m_{e^{\star}}$ at the PWFALC$\otimes$FCC collider.}

\end{figure}

In order to see effect on the background, we plotted the invariant
mass distributions of both the signal and background after the application
of these discovery cuts. Figure 6 shows these distributions with statistical
errors for three collider options. It is seen that the background
values are below the signal peaks when the distributions are examined.
We have applied an additional cut, to extract the signal, on the $e$$\gamma$
invariant mass system as $m_{e^{\star}}-2\Gamma_{e^{\star}}<m_{e\gamma}<m_{e^{\star}}+2\Gamma_{e^{\star}}$,
using the decay width ($\Gamma$) of the excited electron. For the
calculation of statistical significance (SS) of the expected signal
yield, we have used the formula of

\begin{equation}
SS=\frac{\sigma_{S}}{\sqrt{\sigma_{B}}}\sqrt{L_{int}},
\end{equation}

where $\sigma_{S}$ and $\sigma_{B}$ denote signal and background
cross sections, respectively, and $L_{int}$ is the integrated luminosity
of the collider. We have calculated the discovery ($SS\geq5$) and
observation ($SS$$\geq3)$ mass limits of the excited electrons,
assuming the $f=f'=1$ and $\varLambda=m_{e^{\star}}$. The all results
are reported in Table 2. 

\begin{figure}
\begin{centering}
\includegraphics[scale=0.5]{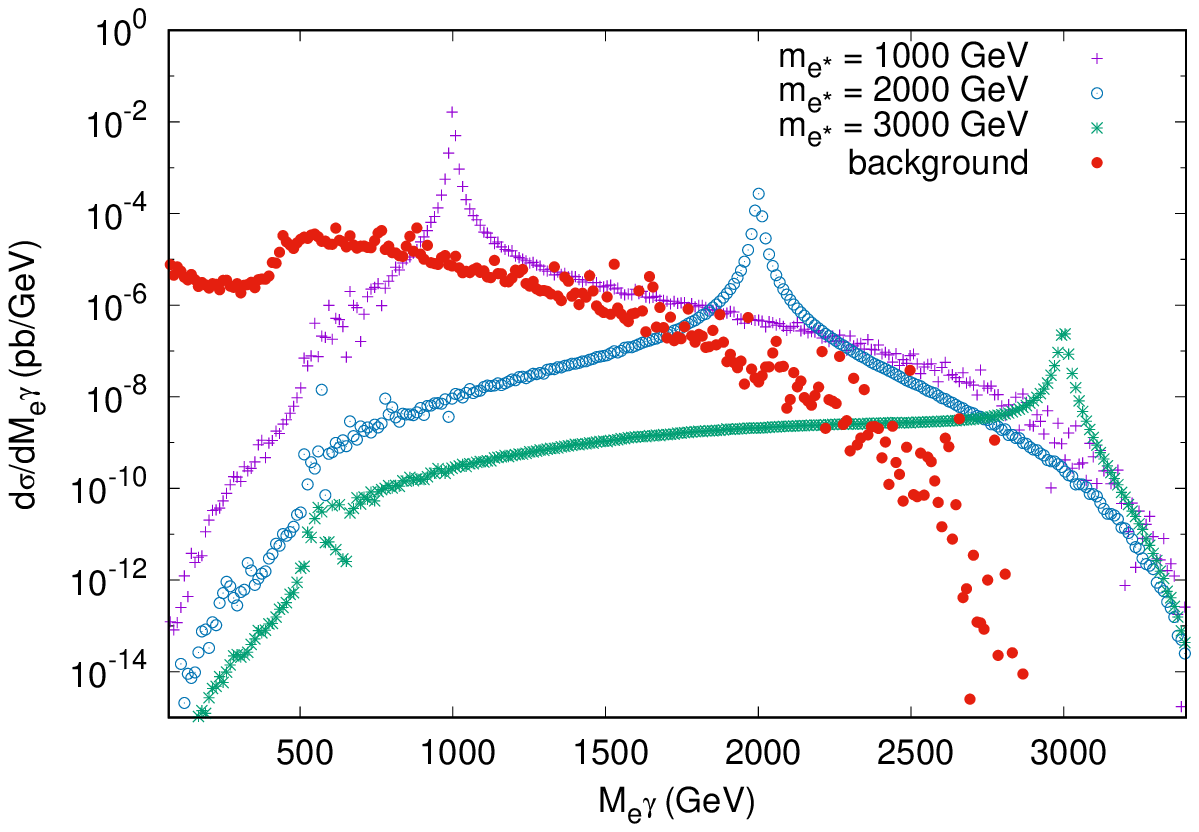}\includegraphics[scale=0.5]{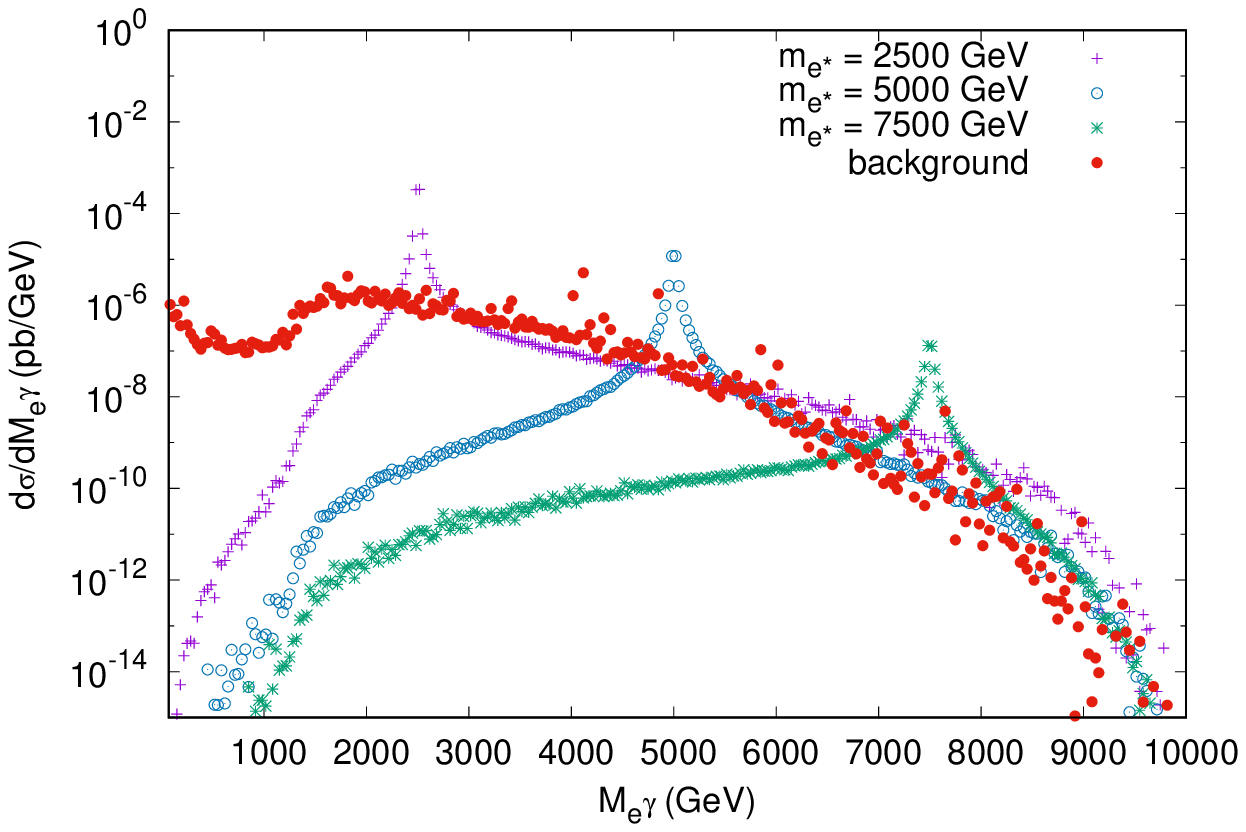}
\par\end{centering}

\begin{centering}
\includegraphics[scale=0.5]{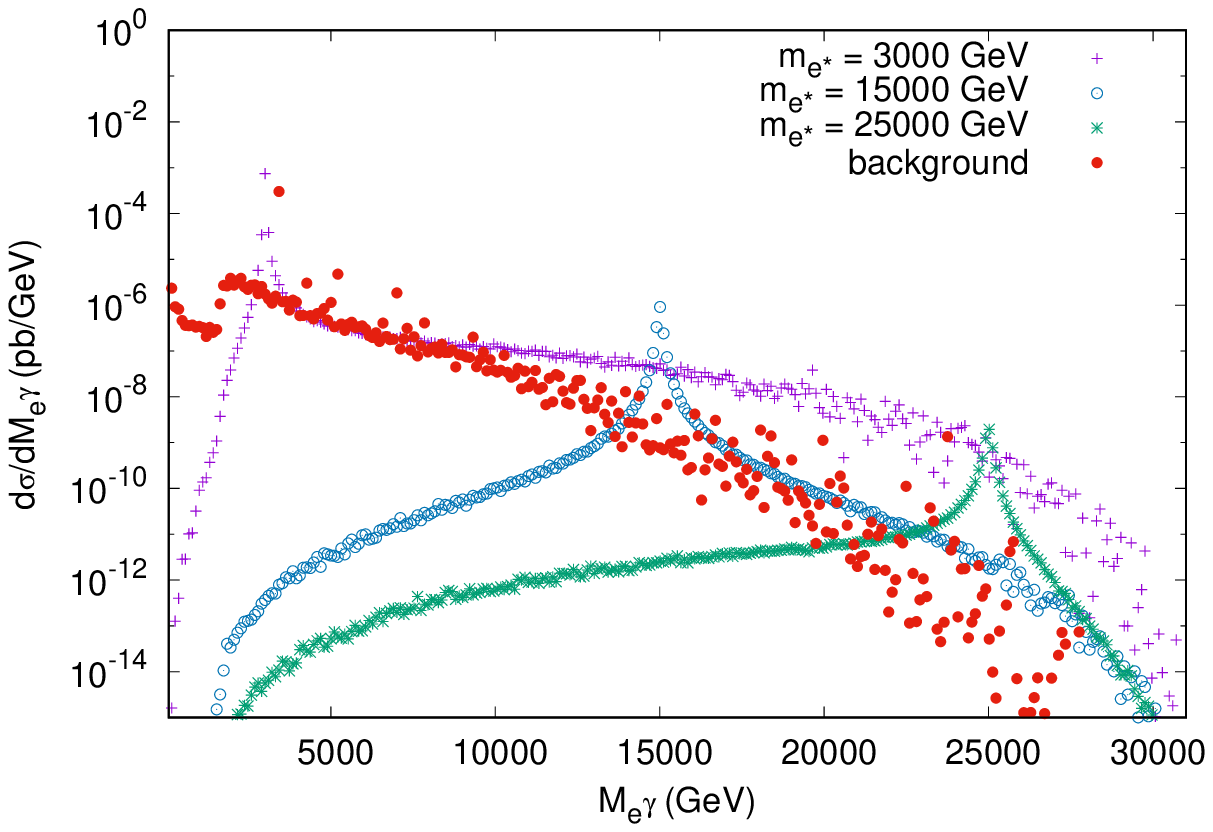}
\par\end{centering}

\caption{The invariant mass distributions, including the statistical errors,
of the electron and photon system for $f=f'=1$ and $\varLambda=m_{e^{\star}}$
at the colliders of ERL60$\otimes$FCC (top-left), ILC$\otimes$FCC
(top-right) and PWFALC$\otimes$FCC (bottom), after the application
of discovery cuts.}

\end{figure}

\begin{table}

\caption{The calculated mass limits for the excited electrons at the FCC-based
electron-proton colliders assuming the coupling $f=f'=1$ .}

\begin{centering}
\begin{tabular}{|c|c|c|c|c|}
\hline 
\multirow{2}{*}{Colliders} & \multirow{2}{*}{$\Lambda$} & \multirow{2}{*}{$L_{int}(fb^{-1})$} & \multicolumn{2}{c|}{$m_{e^{\star}}$$(TeV)$}\tabularnewline
\cline{4-5} 
 &  &  & $3\sigma$ & $5\sigma$\tabularnewline
\hline 
\multirow{2}{*}{ERL60$\otimes$FCC} & $m_{e^{\star}}$ & \multirow{2}{*}{$100$} & $2.4$ & $2.3$\tabularnewline
\cline{2-2} \cline{4-5} 
 & $100TeV$ &  & $2.9$ & $2.7$\tabularnewline
\hline 
\multirow{4}{*}{ILC$\otimes$FCC} & \multirow{2}{*}{$m_{e^{\star}}$} & $10$ & $5.2$ & $4.7$\tabularnewline
\cline{3-5} 
 &  & $100$ & $5.9$ & $5.6$\tabularnewline
\cline{2-5} 
 & \multirow{2}{*}{$100TeV$} & $10$ & $7.9$ & $7.1$\tabularnewline
\cline{3-5} 
 &  & $100$ & $8.3$ & $8.1$\tabularnewline
\hline 
\multirow{4}{*}{PWFALC$\otimes$FCC} & \multirow{2}{*}{$m_{e^{\star}}$} & $1$ & $12.5$ & $11.1$\tabularnewline
\cline{3-5} 
 &  & $10$ & $15.7$ & $14.2$\tabularnewline
\cline{2-5} 
 & \multirow{2}{*}{$100TeV$} & $1$ & $19.7$ & $18.8$\tabularnewline
\cline{3-5} 
 &  & $10$ & $25$ & $22.3$\tabularnewline
\hline 
\end{tabular}
\par\end{centering}

\end{table}

\section{CONCLUSION}

In this paper, the production potential of the excited electrons predicted
by composite models at the three FCC-based electron-proton colliders,
namely ERL60$\otimes$FCC ($\sqrt{s}=3.46$ TeV), the ILC$\otimes$FCC
($\sqrt{s}=10$ TeV) and the PWFALC$\otimes$FCC ($\sqrt{s}=31.6$
TeV), has been investigated. In the analysis made, transverse momentum
and pseudorapidity distributions of the final state particles (electrons
and photons) have been compared for signal and background, and the
appropriate cuts for discovery of the excited electrons have been
determined. And finally, the statistical significance of the expected
signal yield has been calculated and the mass limits of the ecxited
electrons have been assigned.

It is shown that FCC-based electron-proton colliders will be able
to search a very large mass range to detect the excited electrons.
Among them, the collider of PWFALC$\otimes$FCC has the highest center-of-mass
energy. If the excited electrons had not been observed at the ERL60$\otimes$FCC
and the ILC$\otimes$FCC, they would have explored up to the mass
of $22.3$ TeV at the PWFALC$\otimes$FCC collider.

As a result, if the composite electrons exist its probability of being
discovered is high at these FCC-based electron-proton colliders.
\begin{acknowledgments}
We would like to thank Dr. A. Ozansoy for the support of model file.
This work has been supported by the Scientific and Technological Research
Council of Turkey (TUBITAK) under the grant no 114F337.\end{acknowledgments}


\begin{thebibliography}{10}
\bibitem[1]{1}ATLAS Collaboration, ``Observation of a new particle
in the search for the Standard Model Higgs boson with the ATLAS detector
at the LHC'', Phys. Lett. B, 716 (1), 1-29 (2012).

\bibitem[2]{2}CMS Collaboration, ``Observation of a new boson at
a mass of 125 GeV with the CMS experiment at the LHC'', Phys. Lett.
B, 716 (1), 30-61 (2012).

\bibitem[3]{3}I.A. D'Souza and C.S. Kalman, PREONS: Models of leptons,
quarks and gauge bosons as composite objects, World Scientific Publishing,
1992.

\bibitem[4]{4}H. Terazawa, Y. Chikashige and K. Akama, ``Unified
model of the Nambu-Jona-Lasinio type for all elementary-particle forces'',
Phys. Rev. D, 15 (2), 480 (1977).

\bibitem[5]{5}H. Terazawa, ``Subquark model of leptons and quarks'',
Phys. Rev. D, 22 (1), 184 (1980).

\bibitem[6]{6}H. Terazawa, M. Yasue, K. Akama and M. Hayashi, ``Observable
effects of the possible sub-structure of lepton and quarks'', Phys.
Lett. B, 112 (4-5), 387-392 (1982).

\bibitem[7]{7}H. Terazawa, ``A fundamental theory of composite particles
and fields'', Phys. Lett. B, 133 (1-2), 57-60 (1983).

\bibitem[8]{8}H. Fritzsch and G. Mandelbaum, ``Weak interactions
as manifestations of the substructure of leptons and quarks'', Phys.
Lett. B, 102 (5), 319-322 (1981).

\bibitem[9]{9}O.W. Greenberg and J. Sucher, ``A quantum structure
dynamic model of quarks, leptons, weak vector bosons and Higgs mesons'',
Phys. Lett. B, 99 (4), 339-343 (1981).

\bibitem[10]{10}H. Harari, ``A schematic model of quarks and leptons'',
Phys. Lett. B, 86 (1), 83-86 (1979).

\bibitem[11]{11}M.A. Shupe, ``A composite model of leptons and quarks'',
Phys. Lett. B, 86 (1), 87-92 (1979).

\bibitem[12]{12}A. Caliskan, S.O. Kara, A. Ozansoy, ``Excited muon
searches at the FCC-based muon-hadron colliders'', Adv. High Energy
Phys., 2017, 1540243 (2017).

\bibitem[13]{13}A. Caliskan, ``Excited neutrino search potential
of the FCC-based electron-hadron colliders'', Adv. High Energy Phys.,
2017, 4726050 (2017).

\bibitem[14]{14}A. Ozansoy and A.A. Billur, ``Search for excited
electrons through $\gamma\gamma$ scattering'', Phys. Rev. D, 86,055008
(2012).

\bibitem[15]{15}M. Köksal, ``Analysis of excited neutrinos at the
CLIC'', Int. J. Mod. Phys., A29, 1450138 (2014).

\bibitem[16]{16}A. Ozansoy, V. Ari, V. Cetinkaya, ``Search for excited
spin-3/2 neutrinos at LHeC, Adv. High Energy Phys., 2016, 1739027
(2016).

\bibitem[17]{17-1} S. Biondini, O. Panella, G. Pancheri, Y. N. Srivastava,
L. Fano, ``Phenomenology of excited doubly charged heavy leptons
at LHC'', Phys. Rev. D, 85, 095018 (2012).

\bibitem[18]{18-1}R. Leonardi, O. Panella, L. Fano, ``Doubly charged
heavy leptons at LHC via contact interactions'', Phys. Rev. D, 90,
035001 (2014).

\bibitem[19]{19-1}R. Leonardi, L. Alunni, F. Romeo, L. Fano, O. Panella,
``Hunting for heavy composite Majorana neutrinos at the LHC'', Eur.
Phys. J. C., 76, 593 (2016).

\bibitem[20]{20-1}S. Biondini and O. Panella, ``Leptogenesis and
composite heavy neutrinos with gauge-mediated interactions'', Eur.
Phys. J. C., 77, 644 (2017).

\bibitem[21]{21-1}O. Panella, R. Leonardi, G. Pancheri, Y. N. Srivastava,
M. Narain, U. Heintz, ``Production of exotic composite quarks at
the LHC'', Phys. Rev. D, 96, 075034 (2017).

\bibitem[22]{22-1}Y.O. Günayd\i n, M. Sahin and S. Sultansoy, ``Resonance
production of excited u-quark at the FCC based $\gamma p$ colliders'',
e-print, arXiv: 1707.00056 {[}hep-ph{]} (2017).

\bibitem[23]{23-1}H1 Collaboration, ``Search for excited electrons
in ep collisions at HERA'', Phys. Lett. B, 666, 2 (2008).

\bibitem[24]{24-1}D0 Collaboration, ``Search for excited electrons
in $p\bar{p}$ collision at $\sqrt{s}=1.96$ TeV'', Phys. Rev. D,
77, 091102, (2008).

\bibitem[25]{25-1}ATLAS Collaboration, ``Search for excited electrons
and muons $\sqrt{s}=8$ TeV proton-proton collisions with the ATLAS
detector'', New J. Phys., 15, 093011 (2013).

\bibitem[26]{26-1}CMS Collaboration, ``Search for excited leptons
in proton-proton collisions at $\sqrt{s}=8$ TeV'', JHEP, 2016, 125
(2016).

\bibitem[27]{27-1}C. Patrignani et al., (Particle Data Group), ``Review
of particle physics'', Chin. Phys. C, 40, 100001 (2016).

\bibitem[28]{28-1}FCC Poject Web Page: https://fcc.web.cern.ch.

\bibitem[29]{29-1}TLEP Project Web Page: https://tlep.web.cern.ch.

\bibitem[30]{30-1}Y. C. Acar et al., ``Future Circular Collider
based lepton-hadron and photon-hadron colliders: luminosity and physics'',
Nucl. Instrum. Meth., A871, 47-53 (2017).

\bibitem[31]{31-1}C. Adolphsen et al., ``The International Linear
Collider Technical Design Report - Volume 3.2'', e-print, arXiv:1306.6328
{[}physics.acc-ph{]} (2013).

\bibitem[32]{32-1}J. P. Delahaye et al., ``A beam driven plasma-wakefield
linear collider from Higgs factory to multi-TeV'', Proceedings of
IPAC2014, Dresden, Germany, 3791 (2014).

\bibitem[33]{33-1}LHeC Project Web Page: http://lhec.web.cern.ch.

\bibitem[34]{34-1}F. Zimmermann et al., ``Challenges for highest
energy circular colliders'', Proceedings of IPAC2014, Dresden, Germany,
1-6 (2014).

\bibitem[35]{35-1}K. Hagiwara, D. Zeppenfeld and S. Komamiya, ``Excited
lepton production at LEP and HERA'', Z. Phys. C, 29, 115 (1985).

\bibitem[36]{36}U. Baur, M. Spira and P. M. Zerwas, ``Excited-quark
and -lepton production at hadron colliders'', Phys. Rev. D, 42, 815
(1990).

\bibitem[37]{37}F. Boudjema and A. Djouadi, ``Looking for the LEP
at LEP. The excited neutrino scenario'', Phys. Lett. B, 240, 485-491
(1990).

\bibitem[38]{38}F. Boudjema, A. Djouadi and J. L. Kneur, ``Excited
fermions at $e^{+}e^{-}$ and $ep$ colliders'', Z. Phys. C, 57,
425 (1993).

\bibitem[39]{39}A. Belyayev, N. D. Christensen and A. Pukhov, ``CalcHEP
3.4 for collider physics within and beyond the Standard Model'',
Comput. Phys. Commun., 184, 1729 (2013).

\bibitem[40]{40}D. Stump et al., ``Inclusive jet production, parton
distributions and the search for new physics'', JHEP, 0310, 046 (2003).\end{thebibliography}
\end{document}